\documentclass[preprint,showpacs,preprintnumbers,amsmath,amssymb,prb]{revtex4}
\usepackage{graphicx}
\usepackage{dcolumn}
\usepackage{bm}

\usepackage{fancyhdr}
\usepackage{latexsym}

\begin{document}

\title{Introducing  Relativity in Global Navigation Satellite
Systems}

\author{ J.-F. Pascual-S\'anchez}
\affiliation{Dept. Matem\'atica Aplicada, Facultad de Ciencias,
 Universidad de Valladolid, Valladolid, 47005, Spain, EU.}

 \email{e-mail: jfpascua@maf.uva.es}

\begin{abstract}
 Today, the Global Navigation Satellite
Systems, used as global positioning systems, are the GPS and the
GLONASS. They are based on a Newtonian model and hence they are
only operative when several relativistic effects are taken into
account. The most important relativistic effects (to order
$1/c^2$) are: the Einstein   gravitational blue shift effect of
the satellite clock frequency (Equivalence Principle of General
Relativity) and the Doppler red shift of second order, due to the
motion of the satellite (Special Relativity).
 On the other hand, in a few years the Galileo system
will be built, copying the GPS system unless an alternative
project is designed. In this work, it will be also shown that the
SYPOR project, using fully relativistic concepts,  is an
alternative to a mere copy of the GPS system. According to this
project, the Galileo system would be exact and there would be no
need for relativistic corrections.
\end{abstract}

\maketitle

\section{Introduction}
The general purpose of this work is to give  a  very simple, yet
rigorous, mathematical derivation of the relativistic effects on
Global Navigation Satellite Systems (hereafter GNSS),
  suitable as an introduction to the subject.
Also, I will focus on the essential differences between a
Newtonian
 plus relativistic corrections framework, as that of the current
 operating systems, and a fully relativistic framework which would
 be desirable to implement in the future Galileo system due
  to its theoretical and practical advantages.

  At present, all the GNSS,  functioning only as  global positioning systems,
are the GPS (Navstar) (USA) \cite{gps} and the GLONASS (Russia)
\cite{glo}. Both started  as military oriented networks. Nowadays,
there are many important civilian applications of the GNSS such
as: positioning and timing; comparison of time standards around
the world; movements of buildings and bridges that can be
monitored in real time; measurements of relative movements of the
parts of the Earth's crust to forecast earthquakes; location of
missing persons; tests of Special and General Relativity, etc. As
one can see, the civilian applications of the GNSS have overtaken
military applications.

The main characteristics of the present GNSS  will be described in
Sec. II. They are based on a Newtonian model which is corrected
numerically by numerous and important ``relativistic effects".
This Newtonian model uses  a non-inertial  spatial reference frame
for navigation that co-rotates with the Earth and is geocentric:
the ECEF (Earth Centered Earth Fixed system). In the GPS, the ECEF
is
 the military reference system World Geodetic System-1984, WGS-84.
 Moreover, this Newtonian model uses as
 time reference, the coordinate time of a local inertial
 ``star-fixed" reference frame with origin in the Earth's center of mass
 and freely falling with it,
the ECI (Earth Centered Inertial system). An introduction to the
navigation equations will be given in Sec. III.

In general, the satellites of the GNSS are affected by Relativity
in three different ways: in the equations of motion, in the signal
propagation and in the beat rate of the satellite clocks. In this
work I will focus  on the clock effects because they are  the only
measurable ones in the present GNSS. Among the relativistic
effects on the rate of clocks, the most important ones (to order
$1/c^2$) are: the Einstein effect or gravitational frequency shift
of the satellite atomic clocks  (Equivalence Principle of General
Relativity) and the Doppler shift of second order, due to the
motion of the satellites (Special Relativity). If they were not
corrected, the system would not be operative after a few minutes.
In a day of operation, they would produce an error of more than 11
km in the horizontal positioning of the receiver. In a  week, the
error accumulated in the vertical positioning would be of  5 km
approximately.

 I will describe these
effects in Sec. IV  by means of the first post-galilean
approximation to Special Relativity and the first post-Newtonian
approximation to General Relativity using the static spherically
symmetric Schwarzschild's metric
 (or even using the spacetime
  metric induced by the Equivalence Principle in the particular
  case of  circular orbits) for the Earth.

Nevertheless, a constellation of satellites equipped with  clocks
broadcasting  their proper times by means of  electromagnetic
signals, is indeed a proper relativistic or Einstein system.
According to the SYPOR (a French acronym for  SYst\`eme de
POsitionnement Relativiste) project for the Galileo (EU) by  Coll
 \cite {coll2, coll3, coll4}, which is a fully
relativistic way to understand a GNSS system, the Control Segment
should be in the constellation of satellites and therefore
coincide with the Space Segment and not  in the terrestrial geoid
as in the current systems. Thus,   the usual reading would be
inverted and the function of the Control Segment would not be to
determine the position of the satellites or space vehicles
(hereafter SVs) with regard to certain terrestrial coordinates,
but to define these with respect to the constellation of
satellites.

According to the SYPOR project the Galileo system would be exact
and there would be no  need for ``relativistic" corrections. The
main ideas behind this project will be discussed in the final Sec.
V.

 Globally, the current situation
in the GNSS is almost analogous to the following one: imagine that
a century after Kepler, the astronomers were still using Kepler's
laws  as  algorithms to correct the Ptolemaic epicycles  by means
of ``Keplerian effects''. Similarly, a century after Einstein, one
still uses the Newtonian theory and corrects it by ``relativistic
or Einstenian effects'' instead of starting with Einstein's
gravitational theory right away.

 \section{Main characteristics of the GPS system}

  Here we consider only the GPS system because the GLONASS is very similar.
  As it is well known,  see \cite{gps,dana}, the GPS system is formed
  by three different parts  or segments (see Fig. 1):

 The nominal GPS operational constellation
 consists of 24 SVs that orbit the Earth in 12 sidereal hours.
 There are often more than 24 operational satellites as new ones are
 launched to replace older satellites.
 Nowadays, for instance, the current GPS constellation consists of 29
  satellites.

 The orbit altitude is 20,183 km,
  such that the satellites repeat
   the same track and configuration over any point (as the earth turns beneath them)
   approximately every
   24 hours (4 minutes earlier each day). There are six orbital planes
   (with nominally four SVs in each), equally spaced (60 degrees apart),
   and inclined at about fifty-four degrees with respect to the equatorial plane.

    The constellation provides the user or receiver with between five and eight SVs visible
    from any point on the Earth.
    Each satellite contains four atomic clocks,
    two of  Cesium and two of Rubidium  (precision
     of 4 nanoseconds per day), and they emit
      electromagnetic signals that include a code which indicates
      to  the      user its position with respect to the surface of the
      Earth in an ECEF system of reference  and its GPS coordinate time
       which  is advanced with respect to UTC (Universal Time Coordinated).

 {\bf 2) Control Segment}

 The Control Segment consists of four major components:
1.- The Master Control Station (MCS), located at Schriever Air
Force Base (formerly Falcon AFB) in Colorado. 2.- A Backup Master
Control Station, located at Gaithersburg, Maryland. 3.- Four
ground antennas that provide near real-time telemetry, tracking,
and commanding interface among the GPS satellites and the MCS and
4.- Six Monitor quasi-equatorial stations around the world. These
monitor stations measure signals from the SVs which are
incorporated into orbital models for each satellite.

There are models that compute precise orbital data (ephemerides)
and SV clock corrections for each satellite. The Master Control
station uploads ephemerides and clock data to the SVs. The SVs
then send subsets of the orbital ephemerides data to GPS receivers
over radio signals.

 {\bf 3)  User Segment}

The GPS User Segment consists of the GPS receivers and the user
community. GPS receivers convert SVs signals into position,
velocity, and time estimates. Four satellites are required to
compute the four coordinates of X, Y, Z (position) and T (time).

The position is computed  with an error of 10 meters approximately
(it depends on the positions of the SVs for a given location and
time), although with the so-called differential techniques, the
error can decrease to a few meters or even to a lower order. The
idea behind differential positioning techniques (DGPS) is to
correct bias errors at one location with measured bias errors at a
known position. A reference receiver  or base station computes
corrections for each satellite signal. GPS receivers are used for
navigation, positioning, time transfer, and other types of
research.

\section{Four-dimensional Geodesy }

The problem of positioning for a receiver on Earth (E) consists of
calculating his spacetime position $({\mathbf R}_E, T_E)$, from
four spacetime events emitted by four satellites $({\mathbf
r}_{S{_j}}, t_S{_j}), $ $j\,=\, 1,2,3,4$.

The equations of navigation of the GPS are based on the
application of the second principle of Special Relativity
(constancy of the speed of  light in vacuum), in usual Minkowskian
coordinates, they are:
\begin{equation}\label{2}
|{\bf R} _E \, -\,{\bf r}_{S{_j}} |^2 =
c^2\,(T_E\,-\,t_{S{_j}})^2\; ; \,j=1,2,3,4\,.
\end{equation}
These are the equations of  four  light rays (electromagnetic
signals) that travel on four light cones   (3-dimensional
hypersurfaces) from the satellites to the receiver in Minkowski's
spacetime and
 are only valid in a local inertial reference frame in the presence of a gravity field.
In the GPS, this reference frame is the Earth Centered local
Inertial (ECI) reference system whose origin coincides with the
Earth's center of mass. It is in free fall  and does not turn with
the Earth, i.e., it is non-rotating. This local inertial frame,
ECI, is related by a coordinate transformation to the
International Celestial Reference Frame (ICRF). The ICRF is a
local inertial frame defined by the coordinates of 717
extragalactic radio sources and its origin is at the center of
mass of the solar system.

From the point of view of the  Newtonian reference systems, the
first mission of the GPS is to determine the 3-dimensional
position of the receiver on Earth (that is east-west, north-south
and vertical) in the ECEF reference system. In principle, the
signals from three satellites  provide this information. Each
satellite sends a signal that  codifies where the satellite is and
the time of emission of the signal. The  receiver clock times the
reception of each signal, then subtracts the time of emission to
determine the lapse of time travel  and hence the distance that
the signal has travelled from the SVs at the speed of light.

This is the distance   the satellite  was from the receiver when
it emitted the signal. Taking these three distances as radii,
three spheres $S^2$ are constructed with center on each satellite.
The receiver is located in one of the two intersection points  of
the three spheres $S^2$. This is, in short, the procedure of
trilateration.

Nevertheless, there is still a problem: the  receiver clock is
usually made of quartz crystals oscillators and hence it is not as
accurate as the atomic clocks in the SVs. For this reason, the
signal of a fourth satellite is needed to obtain the error in the
precision of the receiver clock and  enables the  user's clock to
be used as  an atomic clock.

We will see below that, from the point of view of the relativistic
positioning systems, the receiver determines its four coordinates
(in general since we will see, not necessarily three of coordinate
space and one of coordinate time, but four proper times),
identifying the points of intersection of the four light cones of
the satellites that intersect  the light cone of a receiver, see
Fig. 2.

\section{Relativistic effects in the GPS}
In the GPS constellation, the satellite clocks keep time to an
accuracy of about four nanoseconds per day ($4\, {\rm ns/d} \, = 4
\times 10 ^ {-9} \, {\rm s/d} \, $), which  amounts to a
fractional time stability of 5 parts in $10^{14}$. This means that
once an atomic clock of in the SVs is set, then after one day it
should be correct to within
 about 4 nanoseconds. Therefore, the errors introduced by the relativistic effects at
order $c^{-2}$ were crucial in its conception, because these
effects  are of the order of hundred or thousand nanoseconds. To
ignore  these effects would make the GPS useless for three main
reasons \cite{pas1}:

{\bf First. Einstein gravitational blue shift effect}.

 The clocks  run faster when they are far away from a center of
gravitational attraction. This center is the center  of the Earth
in the case of GPS. Hence, there is a blue shift of the frequency
for a signal sent downward from the satellites to the Earth, due
to usual Newtonian gravitational effects depending on the distance
to the center of the Earth. We stress that, in a time independent
gravitational field (as the gravitational field of the Earth in
first approximation) the coordinate frequency of the radio signal
is conserved as it propagates, hence the blue shift of the photon
at its reception on Earth is due to the fact that the time scale
is slower for the Earth bound clock.

{\bf Second. Time dilation or special relativistic Doppler effect
of second order}.

The motion of both the SVs and  the Earth
receivers must be taken into account. As the satellite clocks move
faster than the clocks on the Earth's surface this gives rise, on
the one hand  to the classical Doppler effect  (linear in speed),
and to the other to the special relativistic Doppler effect of
second order in the satellite
 speed $O({v^2_{_{S}}}/{c^2})$ due to time dilation
 (clocks  in motion run slow). Consequently, the latter gives place to
 a red shift of the frequency for a signal sent downward from the
satellites to the Earth.

{\bf Third. Sagnac  effect and gravitomagnetic field of rotation}.

In principle, it is necessary to consider the Earth rotation for
two reasons. Firstly,  the  kinematical Sagnac effect \cite{pas,
ash04}, which is of order $c^{-2}$  and produces a maximum error
of the order of $207.4\, {\rm ns/d}$ in the GPS. Secondly, the
effect of the gravitomagnetic field (GM) \cite{mas}
  generated by the rotation of the Earth's mass. Even today
the effects of this GM field  have not been incorporated to GNSS
because they are of order $O(c^{-3})$, i.e. picoseconds, and hence
negligible at the accuracy presently required.

 The Schwarzschild metric appears to be  sufficient to correct
  the relativistic effects at order $c^{-2}$,
 something which is necessary to describe the spacetime near the
Earth. Nevertheless, strictly speaking, the static metric with
spherical symmetry of Schwarzschild does not describe exactly the
spacetime near the Earth for two reasons:

  1) The Earth is not spherical.

  2) The Earth's mass does not remain static but, when turning
   it generates a stationary field of rotational gravity
(non-Newtonian) or gravitomagnetic field (GM). Thus, the external
gravity field of the Earth can be described  more adequately by a
Kerr metric or its Lense-Thirring linear approximation.

   However,  the Earth rotates slowly with an angular velocity at the equator of
 $\omega_{_{E}} =  7.2921151247 \times  10^{-5} \;{\rm rad\; s^{-1}}$
 or, equivalently, at a  linear speed of $v _{_{E}}= 465 \;{\rm m \;
 s^{-1}}$.
  Hence so far, with the required precision in the measurement of
the time (ns),  the Schwarzschild metric is a sufficient
approximation  for the objetives of the GNSS.

Let us remark nevertheless, that according to the  IAU-2000
resolutions the gravitomagnetic effects, which have the order
$O(c^{-3})$,  must be incorporated to define as the barycentric
celestial reference system  (BCRS) as well as the geocentric
celestial reference system (GCRS). So, in Astrometry the
gravitomagnetic effects must be  considered because the order of
accuracy is greater than in the current GNSS.

 In  classical Droste-Hilbert coordinates the expression
 of the Schwarzschild metric for equatorial
orbits, reads as:
\begin{equation}\label{1} ds^2 = c^2 \,d\tau^2=
 \left(1-\frac{2GM}{rc^2}\right)\,c^2\,dt^2 -  \left
(1-\frac{2GM}{ r\,c^2}\right)^{-1} dr^2 -r^2 d\psi^2~,
\end{equation}
where $\tau$ is the  local proper time of a comoving observer at
rest, $t$ is far-away  global  time or coordinate time measured in
an  inertial frame placed at spatial infinity  and $\psi$ is the
azimuthal angle.

   Let us consider a circular orbit (see note \cite{comment}), then the clock on the
Earth and the satellite clock travel at constant distance around
the Earth center, therefore $dr = 0$ for each clock. For both
clocks, one obtains:
\begin{equation}\label{3}
\frac{1}{c^2}\left(\frac{ds}{dt}\right)^2 =
\left(1+\frac{2\,\phi}{ c^2 }\right)\,-\,\frac{v^2}{c^2}\,,
\end{equation}
where $\phi$ is the   Newtonian potential  and
   $v = r \frac{d\psi}{dt}$ is the coordinate
 tangential speed along the circular
equatorial orbit, measured using the coordinate far-away time $t$.

Now, let us apply  Eq. (\ref{3}) twice, first
  to the clock in a satellite (S), (using $r = r_{_{S}}$, $v = v_{_{S}}$ and proper
time $d\tau_{_{S}}= ds/c$) and secondly to a fixed clock in the
Earth's equator and turning with it (E) (using, $r = r_{_{E}}$, $v
= v _{_{E}}$ and  proper time $d\tau_{_{E}}=ds/c$), with the same
 elapsed coordinate time, $dt$, corresponding to an inertial observer at spatial
infinity. Then,  dividing both expressions one obtains:
\begin{equation}\label{4}
\left(\frac{d\tau_{_E}}{d\tau_{_S}}\right)^2=
\frac{1-\displaystyle\frac{2\,GM}{r_{_{E}}\,c^2}\,-\,\frac{v^2_{_{E}}}{c^2}
}{1-\displaystyle\frac{2\,GM}{r_{_{S}}\,c^2}\,-\,\frac{v^2_{_{S}}}{c^2}}.
\end{equation}
The relativistic effects we are considering  are only those of
order $O(c^{-2})$. This is the order of approximation used in the
GPS. So, why is this order of approximation  good enough? Because,
as  said above, the accuracy of the SVs atomic clocks is
nanoseconds and only these three effects are of higher magnitude.

{\sl Are the   relativistic effects considered  small at this
order? No,} as we will show, they are not only important, but
crucial.

On the other hand, it is important to point out that the necessity
to consider  other smaller relativistic effects
 at the order $O(c^{-3})$  when laser cooled atomic clocks
   of last  generation   are
used, with measurable errors of  order of picoseconds, as in the
ACES (Atomic Clock Ensemble in Space) mission of the ESA for the
ISS \cite{blanchet1}, and those of the order $O(c^{-4})$
\cite{LIN,TEYS},  when one
 works  with intrinsic errors of femtoseconds in the near future.

 \subsection{Einstein gravitational blue shift effect:\,45,700\,ns/d}
 Now, let us suppose that we ignore the motions of the satellite clock
 and of the clock on the Earth's surface, i.e, let us start
ignoring the Doppler effect of second order of Special Relativity,
that is to say both are fixed in the ECI:  $v _{_{E}}= v_{_{S}}=
0$.
 Then,  Eq. (\ref{4}) reduces to
 \begin{equation}\label{6}
 \frac{d\tau_{_E}}{d\tau_{_S}}=
\frac{\left(1-\displaystyle\frac{2\,GM}{r_{_{E}}\,c^2}
\right)^\frac{1}{2}}{\left(1-\displaystyle\frac{2\,GM}{r_{_{S}}\,c^2}\right)^\frac{1}{2}}\,.
 \end{equation}
 The radius of a circular orbit of a SV with a period of 12
 sidereal hours
  is approximately $r_{_{S}}= 26,561 $ km.
 The linear approximation of (\ref{6}) reads as:
\begin{equation}\label{8}
\frac{d\tau_{_E}}{d\tau_{_S}}=
1-\displaystyle\frac{GM}{r_{_{E}}\,c^2}
+\frac{\,GM}{r_{_{S}}\,c^2}  =1- \,D\,.
\end{equation}
  The corresponding numerical values for these terms are:
\begin{equation}\label{9}
 \displaystyle\frac{GM}{r_{_{E}}\,c^2}=
6.95\times10^{-10},
\end{equation}
and
\begin{equation}\label{10}
 \displaystyle\frac{GM}{r_{_{S}}\,c^2}=
1.67\times10^{-10}.
\end{equation}

In Eq. (\ref{8}) $D$ is a  positive number   and gives an estimate
of the Einstein gravitational blue shift, between the stationary
clocks in the position of satellite and on the Earth's surface. A
displacement towards the blue is observed in the clocks on the
Earth  when a radio signal is sent downward from the satellites to
Earth.

 Is this  difference negligible  or is it crucial for the functioning
of GPS?. Taking into account that in one nanosecond an
electromagnetic signal travels about
  30 cm, we have that if a difference of hundreds of nanoseconds
  causes difficulties in the accuracy of the GPS, then  a difference of
thousands of nanoseconds like that which  is obtained with the
Einstein effect will cause even greater difficulties in the accuracy.\\
 The clock in the satellite runs faster, i.e., it goes ahead about
 45.7 ${\rm \mu s/d}$ compared with a clock on the Earth's surface, due to
gravitational effects associated with position exclusively.
General Relativity is clearly needed  for the correct operation of
the GPS.

\subsection{Time dilation or special relativistic Doppler
effect of second order: \,7,100\,\,ns/d} Besides the gravitational
effects  of the position of the satellite and the observer on the
Earth, we must  also consider the effects of Special Relativity
due to relative motion of the clocks in the satellite and of a
clock  fixed on the  Earth's surface. The clock of the satellite
moves at greater speed than a fixed clock on the Earth's surface.
Their speeds are,
 $v_{_{S}}= \,3.874\;{\rm km/s}$\,, and \,
 \,$v _{_{E}}= 465 \;{\rm m/s}$, respectively.
 It is well known that Special Relativity predicts  that  ``clocks in motion slow down"
  or, in other words , to an SV higher speed  corresponds
   a  red shift when the light is received on Earth.

Therefore, {\sl this kinematic Doppler effect  of second order
 ``works against"  the Einstein gravitational blue shift effect}. At the critical altitude
 of $H= \,3,167 \,{\rm km}$ they  compensate each other, but at the altitude
  of the satellites of the GPS, $H_{GPS} = \,
 20,183 \, \,{\rm km}$, the Einstein  gravitational blue shift effect  is
 dominant.

 The final formula that accounts for both effects is obtained
 from Eq. (\ref{4}) which in the first order of approximation  reads:
\begin{equation}\label{12}
\frac{d\tau_{_E}}{d\tau_{_S}}=
1-\displaystyle\frac{GM}{r_{_{E}}\,c^2}\,-\,\frac{v^2_{_{E}}}{2\,c^2}
+\displaystyle\frac{GM}{r_{_{S}}\,c^2}\,+\,\frac{v^2_{_{S}}}{2\,c^2}\,.
\end{equation}

An analogous formula (except for a change of sign) for circular
equatorial orbits, was first given in \cite{sin}. Introducing
numerical values in (\ref{12}), the net result is of the order of
$39,000 \, \, {\rm ns/d}$ , which is equivalent to an error of
11,700 m after a day of operation.

\subsection{GPS coordinate time}
As we will see, the previous net result, in which only proper
times appear, must be corrected using the so-called GPS coordinate
time and a more realistic gravitational potential for the Earth.
We restart with a more realistic metric for the weak external
gravitational field of the Earth which is obtained from the
linearized Schwarzschild metric added to the gravitational
potential of the mass quadrupole moment. The GPS uses this
approximated metric to describe the spacetime near the Earth. This
metric, when expressed in the ECI reads, to order $O(c^{-2})$, as:
\begin{equation}\label{13}
ds^2 =  \left(1+\frac{2V}{c^2}\right)\,c^2\,dt^2 - \left
(1-\frac{2\,V}{c^2}\right) dr ^2 - r^2d\Omega^2~.
\end{equation}
Here   $V$ is  the Newtonian potential $\phi$ plus  the Newtonian
quadrupole potential of the Earth;  and where $d\Omega^2 =
d\theta^2 + {\rm sin}^2\theta\,d\psi^2$.

Nevertheless, the ephemerides of the satellites are calculated in
the ECEF system. Thus, carrying out the transformation of
coordinates from the ECI to the ECEF  ($t = t' \,;\,r= r'\,;\,
\theta = \theta' \,;\, \psi = \psi' + \omega_{_{E}} \,t'$), the
spacetime metric in the ECEF reference frame has the expression:
\begin{eqnarray}\label{14}
ds^2 &=&\left(1+\displaystyle{\frac{2\,\Phi_{\text{eff}}}{c^2}}
\right)c^2\,dt^2 -
2\,\omega_{_{E}} r^2 \sin^2\theta\, d\psi'\,dt\nonumber\\
   & &- \left(1-\displaystyle{\frac{2\,V}{c^2}}\right)
dr ^2 - r^2 d\Omega^2 ~,
\end{eqnarray}

 where the cross term
in Eq. (\ref{14}) gives rise to the kinematical Sagnac effect,
\cite{pas,ash04}, due to the rotation of the ECEF, and where
$\Phi_{\text{eff}}$ is the effective potential in the ECEF
rotating frame, which now contains also the centripetal potential,
and reads:
\begin{equation}\label{eff}
\Phi_{\text{eff}}= V - \displaystyle{\frac{\omega_{_{E}}^2\,r^2
\sin^2\theta}{2}}.
\end{equation}

The equipotential condition $ \Phi_{\text{eff}}= \Phi _ {0}=
const$, defines the  geoid model or reference ellipsoid of
geodesists. The term $\Phi _ {0}/c^2$, when it is evaluated at the
equator ($r=r_{_E}$, $\theta= \pi/2$), has the form:
\begin{equation}\label{0}
\frac{\Phi _ {0}}{c^2}=
-\frac{GM}{r_{_E}c^2}-\frac{J_2GM}{2r_{_E}c^2}-\frac{\omega_{_{E}}^2
r_{_E}^2}{2c^2}= -6.9693\times 10^{-10},
\end{equation}
where  the WGS-84  value of the quadrupole coefficient $J_2$ is
$1.0826300\times10^{-3}$ and $\omega_{_{E}} = 15\,{\rm arcsec/s}$.

 Thus, the metric  in the ECEF frame evaluated at the geoid has the expression:
\begin{eqnarray}\label{ecef}
ds^2 &=&\left(1+\displaystyle{\frac{2\,\Phi_{0}}{c^2}}
\right)c^2\,dt^2 - 2\,\omega_{_{E}} r^2 \sin^2\theta\,
d\psi'\,dt\nonumber\\
& &-\left(1-\displaystyle{\frac{2\,V}{c^2}}\right) dr ^2 -
r^2d\Omega^2 ~.
\end{eqnarray}

Now, notice that in order {\sl to define the speeds that appear in
Eq. (\ref{3}),  the coordinate time $t$ of the Schwarzschild
metric measured in an inertial reference frame at spatial infinity
is used instead of  the coordinate time $t_{\text{\sc gps}}$ of
clocks at rest in the geoid}. The relationship between them, as it
can be seen from Eq. (\ref{ecef}), is:
\begin{equation}\label{11}
dt_{\text{\sc gps}}= \left(1+2\,\Phi_{0}/c^2 \right)^{1/2}\,dt,
\end{equation}
or  to first order
\begin{equation}\label{15}
dt_{\text{\sc gps}}=\left(1+\displaystyle\Phi_{0}/c^2
\right)\,dt\,.
\end{equation}
As it appears in Eq. (\ref{0}), $ \Phi_{0}$ is the effective
potential of gravity  on the geoid that  includes the centripetal
term and the quadrupole potential. Also, we have calculated in Eq.
(\ref{0}) the negative correction,  $\Phi_{0}/c^2$, which is of
the order of seven   parts in $10^{-10}$. Thus,  the GPS
coordinate time $t_{\text{\sc gps}}$,   runs   slow with respect
to the coordinate time $t$ measured by an inertial observer at
infinity.

This metric in the ECI local inertial reference frame, when the
relation Eq. (\ref{11}) is introduced in Eq. (\ref{13}), reads as:
\begin{equation}\label{hola}
ds^2 =  \left[1+\frac{2(V - \Phi_{0})}{c^2}\right]c^2dt_{\text{\sc
gps}}^2 - \left (1-\frac{2\,V}{c^2}\right) dr ^2 - r^2 d\Omega^2
~.
\end{equation}
As it can be seen from  Eq. (\ref{ecef}), the global GPS
coordinate time $t_{\text{\sc gps}}$ is, also, the proper time of
the clocks at rest on the geoid where $ \Phi_{\text{eff}}=\,\Phi _
{0}$. It is a time referred to ECEF  (advanced with respect to the
UTC), but coincides with a fictitious time of a clock at rest in
the ECI, see also \cite{lam}. This one is a local inertial
reference frame in which  the navigation equations are applicable
and the Poincar\'e-Einstein synchronization can be established
globally within it \cite{minguzzi}.

 The spacetime metric referred to the ECI, and given in  (\ref{hola})
  can be transformed into:

\begin{eqnarray}\label{18}
ds^2 =& \left[1+\displaystyle\frac{2(V - \Phi_{0})}{c^2} -
\left(1-\displaystyle\frac{2V}{c^2}\right)
  \displaystyle\frac{1}{c^2}\left(\frac{dr}{dt_{\text{\sc
  gps}}}\right)^2
     -\displaystyle\frac{r^2}{c^2}\left(\frac{ d\Omega}{dt_{\text{\sc gps}}}
   \right)^2\right] c^2 dt^2_{\text{\sc gps}}.
\end{eqnarray}

The tangential speed of the satellite in the ECI (supposing
circular orbits, $dr=0$) is:
\begin{equation}\label{19}
 v{_{_S}} =   r \frac{d\Omega}{dt_{\text{\sc gps}}}\,,
\end{equation}
hence the proper time in the satellite, at order $ O(c^{-2})$, is
\begin{equation}\label{17}
d\tau_{_S}= ds/c = \left[1+\frac{(V\, - \,\Phi_{0})}{c^2} -
\frac{v{_{_S}}^2}{2c^2}\right] \,dt_{\text{\sc gps}}\,,
\end{equation}
and, integrating along a trajectory $C$, the GPS coordinate time
reads as:
\begin{equation}\label{20}
\int_{C} dt_{\text{\sc gps}}
 = \int_{C} d\tau_{_S} \,\left[1-\frac{(V\, -
\,\Phi_{0})}{c^2} + \frac{v{_{_S}}^2}{2c^2}\right]\,.
\end{equation}

In this formula the five main sources of error appear:

1º)\,In $ \Phi_{0}$.   Three effects on the fixed clocks in the
ECEF: monopole (the main term responsible of the global blue
shift), terrestrial quadrupole potential and centripetal potential
due to the Earth's rotation, $v^2_{_{E}}/2\,c^2$, (Doppler of
second order effect, blue shift).

 2º)\,In $V$.  The Einstein gravitational effect  due to the
position of the satellite (redshift) and the terrestrial
quadrupole potential.

 3ª)\,In $v{_{_S}}^2/2c^2$. The
special relativistic effect (Doppler of second order) due to the
orbital motion of the satellite (red shift).

Neglecting the Earth's quadrupole moment in $V$ (i.e. supposing
Keplerian orbits, $V=\phi$) and assuming that the SVs orbits are
circular ($v_{_S}^2=GM/r$), one obtains:

\begin{equation}\label{final}
\int_{C} dt_{\text{\sc gps}}
 = \int_{C} d\tau_{_{S}} \,\left(1+\frac{3GM}{2rc^2}\, +\frac{\Phi_{0}}{c^2} \right)\,.
\end{equation}

Numerically, the constant rate correction is:
\begin{equation}\label{num}
\frac{3GM}{2rc^2}\, +\frac{\Phi_{0}}{c^2}=  - 4.4647 \times
10^{-10}= -38.58\,{\rm \mu s/d}.
\end{equation}
The negative sign of the sum of the second and third terms  in
(\ref{final})  means that a clock in orbit gains time relative to
a clock on the Earth. Therefore, {\sl using the global GPS
coordinate time $t_{\text{\sc gps}}$, the clock in a satellite
goes ahead with respect to an Earth fixed clock and the final net
time gained is
 of the order of $38,580 \,{\rm ns/d}$, that is to say an error of
  $11,574$ {\rm m/d}.} In fact, when the first GPS satellite equipped with a Cesium clock
   was   launched in 1977, its relativistic correction program was not activated for 20 days,
    during that time its atomic clock provided a great test for relativity
     losing the  figure that appears in (\ref{num}).

So, for the  satellite atomic clocks  to appear to the observer in
the ECEF at the operating frequency, $10.23 \,{\rm MHz}$, the
atomic clocks of the satellites must be fitted before their
launch, lowering their proper frequency to:
\[\left[1 - 4.4647 \times 10^{-10} \right]\times 10.23 \,\,{\rm
MHz} = 10.229\, 999\, 995\, 43\,\, {\rm MHz}.\] In this way, the
orbiting clock becomes a global coordinate GPS clock and the local
proper time of satellite clocks is not used.

In addition to the above mentioned effects, it is necessary to
correct the path dependent kinematical Sagnac effect or lack of
global synchronization (maximum error of 207.4 {\rm  ns}) due to
the rotation of the ECEF and the effect of eccentricity (because
the orbits are not circumferences but ellipses). In the GPS
(GLONASS), these  two small  effects of order $O(c^{-2})$ are
corrected by the receiver clock on the Earth (clock in the
satellite). It is remarkable that one can use the GPS to
 measure the Sagnac effect as it was done in \cite{allan}.

Other effects of order $c^{-2}$, as the secondary Sagnac effect
due to the polar motion or the tidal effects of the Moon and Sun,
are negligible in the GPS and GLONASS because their  magnitude is
in the order  of picoseconds .

Finally it is necessary to indicate that, when transferring
frequencies, the classical longitudinal Doppler effect (linear in
speed), also appears which in principle must be accounted for by
the receivers. Depending on the speed of the receiver,  it is
$10^{3}-10^{5}$ higher than the secular relativistic effects
considered so far.

A more detailed study of relativistic effects  of order $1/c^2$ on
the GPS can  be found in  \cite{ashby02,ash,bahder01,bahder02};
the  post-Newtonian effects of  order $1/c^3$, as for instance the
Shapiro time delay, in  \cite{blanchet1}, and those of order
$1/c^4$ in \cite{LIN}. Finally, in  \cite{TEYS},
 is  considered the post-Minkowskian approximation  to order
$G^2$, which contains the post-Newtonian effect of order $1/c^5$.

\section{ The SYPOR project  for the Galileo}
\subsection{Basic idea}
  The basic idea of this project which
is expected to be accepted in a Framework Programme of the EU, as
it is was reported in \cite{coll2} and \cite{coll3}, is the
following one: A constellation of SVs with clocks that interchange
their proper time among themselves and with Earth receivers, is  a
fully relativistic system. {\sl In the SYPOR, the Segment of
Control is in the constellation of satellites}, see Fig. 3.

The function of this new Control Segment is not to determine the
ephemerides of the satellites with respect to geocentric
coordinates as in the Newtonian GNSS, but to determine the
coordinates of the receivers with respect to the constellation of
SVs. Therefore,  the procedure used up to now in the Newtonian
GNSS is inverted. The  SYPOR would be composed of two sub-systems:
An independent one, made up of  the constellation of satellites
(primary 4-dimensional positioning system) and another one,
coupling the Earth to the first sub-system (terrestrial secondary
3-dimensional reference system, WGS84 or ITRF and a scale of
time).

 The main technical characteristics of the SYPOR
are: 1) External control of  the {\sl full} system: it consists of
a device in at least four of the SVs pointing to the ICRS
(International Celestial Reference System). 2) Internal control of
the {\sl parts} of the system: a device in each satellite that
exchanges its proper time with other SVs. 3) Control by the
Segment of receivers: it consists of a device in each satellite
that  sends the proper times of its nearby satellites as well as
its proper time to the Earth by means of electromagnetic signals.

The SYPOR is a long term project  which is still in a state of
theoretical construction. We will now summarize its main
conceptual basis. In Relativity, in the presence of gravity, the
spacetime is modelled by a 4-dimensional Lorentzian manifold with
a pseudo-Riemannian metric. So, for a domain of the manifold,
there exists a large variety of mathematical coordinate systems
but only a few number of them admit a physical interpretation and
can be physically constructed by means of a protocol on a set of
physical fields (point test particles, light rays, etc.).

\subsection{Reference systems and positioning systems}
Let us define location systems as the physical realizations of
some coordinate systems. Location systems are of two different
types: reference systems and positioning systems. The first ones
are 4-dimensional reference systems which allow one observer,
considered at the origin, to assign four coordinates to the events
of his neighborhood  by means of  a transmission of information.
Due to the finite speed of light this assignment is intrinsically
retarded with a time delay. The second one are 4-dimensional
positioning systems (as  used in the SYPOR) which allow to every
event of a given domain to know its proper coordinates without
delay or in an immediate or instantaneous way.

In Relativity, a (retarded) reference system can be constructed
starting from an (immediate) positioning system (it is sufficient
that each event sends its coordinates to the observer at the
origin) but not the other way around. In contrast, in Newtonian
theory, 3-dimensional reference and positioning systems are
interchangeable and as the velocity of transmission of
information, the speed of light, is supposed to be infinite, the
Newtonian reference systems are not retarded but immediate.

The reference and positioning systems defined here are
4-dimensional objects, including time location. This is not the
common use, and so, the International Astronomical Union (IAU)
still considers separately time scales and 3-dimensional reference
systems such as the International Celestial Reference System
(ICRS) or the International Terrestrial Reference System (ITRF) .

Following \cite{coll1}, the best  way to visualize and
characterize a space–time coordinate system is to start from four
families of coordinate 3-surfaces, then, their mutual
intersections give six families of coordinate 2-surfaces and four
congruences of coordinate lines. Alternatively,  one can use the
related covectors or 1-forms $ \{ {\bm \theta}^i\}, i = 1,2,3,4$,
instead of the 3-surfaces, and the vectors of a coordinate tangent
frame $ \{{\bm e}_i\}, i = 1,2,3,4$,  instead of four congruences
of coordinate lines which are their integral curves. In this way,
for a specific domain of a Lorentzian spacetime, each coordinate
system is fully characterized by its causal class, which is
defined by a set of 14 characters:
\begin{equation}\label{22}
\{{\rm{c_1\, c_2\, c_3\, c_4, C_{12}\, C_{13}\, C_{14}\, C_{23}\,
C_{24}\, C_{34}}},{\it c_1\, c_2\, c_3\, c_4}\},
\end{equation}
being ${\rm{c_i}}$ the Lorentzian causal character of the vector
${\bm{e}_i}$, i.e. if it is spacelike, timelike or lightlike;
${\rm{C_{ij}}}$ the causal character of the adjoint 2-plane
$\{{\bm{e_}i\wedge\,\bm{e_}j\}}$ and, finally, $\it{c_i}$ the
causal character of the covectors ${\bm \theta}^i$ of the dual
coframe,
 ${\bm \theta}^i({\bm{e}_j})=\delta^i_j$. The covector ${\bm \theta}^i$  is
 timelike (resp. spacelike) iff the 3-plane generated by the
 three vectors $\{{\bm{e}_j}\}_{j\neq i}$ is spacelike (resp.
 timelike). This applies for both Newtonian and Lorentzian
 spacetimes. In addition, for the latter, the covector ${\bm \theta}^i$
 is lightlike iff the 3-plane generated by $\{{\bm{e}_j}\}_{j\neq
 i}$ is lightlike or null.

 This new degree of freedom (lightlike) in the causal
 character, which is proper of Lorentzian relativistic spacetimes
 but which does not exist in Newtonian spacetimes,  allows to obtain, after a
  deep algebraic study \cite{coll1}, the following conclusion: {\sl In the 4-dimensional Newtonian
   spacetime there exist four, and only four, causal classes of frames, whereas in the
    relativistic 4-dimensional Lorentzian spacetime,
   due to the freedom commented above, there exists 199,
   and only 199, causal classes of frames.}

As notation for the causal characters, we will use lower case
roman types ({\rm{s,t,l}}) to represent the causal character of
vectors (resp. spacelike, timelike, lightlike), and capital types
(S,T,L) and lower case italic types ({\it s,t,l}) to denote the
causal character of 2-planes and covectors, respectively.

   The ECI and ECEF Newtonian
   coordinate systems used in the GPS belong to two of the four
   causal classes of Newtonian frames,
   $\{{\rm{t\,s\,s\,s,T\,T\,T\,S\,S\,S}}, t\,s\,s\,s  \}$ and
    $\{{\rm{t\,t\,s\,s,T\,T\,T\,T\,T\,S}}, s\,s\,s\,s  \}$,
    respectively. For instance, the ECI  coordinate tangent frame is
    $\{\bm {e}_1=\partial/\partial t,\bm {e}_2=\partial/\partial r,
    \bm {e}_3=\partial/\partial\theta,
    \bm {e}_4=\partial/\partial \psi\}$ and the corresponding coframe is $
    \{{\bm \theta}^1= dt,{\bm\theta}^2=dr,{\bm\theta}^3= d\theta,
    {\bm\theta}^4=d\psi\}$.

In Relativity, a specific causal class, among the 199 ones, can be
assigned to any of the different  coordinate systems used in all
the  solutions of the Einstein equations. However, for the same
coordinate system and the same solution, the causal class can
change depending on the region of the spacetime considered.

 A coordinate system can be constructed starting from a tangent
frame or its coordinate lines. But  in this way, in general, there
are obstructions to obtain   positioning systems that are generic,
i.e., valid for many different spacetimes. On the other hand,
generic positioning systems can be constructed  from four
one-parameter families of 3-surfaces or, equivalently, from a
cotangent frame. One clock broadcasting its proper time is
described in the spacetime by a timelike line in which each event
is the vertex of a future light cone. The set of these light cones
of a emitter constitutes a one-parameter (proper time) family of
null hypersurfaces. So, four clocks broadcasting their proper
times determine four one-parameter families of null or lightlike
3-surfaces.

In a relativistic spacetime, the wave fronts of those signals
parameterized by the proper time of the clocks, define four
families of light cones (3-dimensional null hypersurfaces) which
contain all the lightlike geodesics of four emitters and  making a
contravariant null (or light) coordinate system \cite{coll1}, see
Fig. 2. Such a coordinate system does not exist in a Newtonian
spacetime where the light travels at infinite speed.

\subsection{Coll positioning systems}
 Among the 199 Lorentzian causal classes, only one is privileged
to construct a generic (valid for a wide class of spacetimes),
gravity free (the previous knowledge of the gravitational field is
not necessary) and immediate positioning system.  This is the
causal class $\{{\rm{s\,s\,s\,s,S\,S\,S\,S\,S\,S}}, l\,l\,l\,l \}$
of the Derrick-Coll-Morales coordinate system \cite{derrick,
coll5}. This class includes the {\sl emission coordinates} of the
Coll positioning systems \cite{coll3, coll4}. A coordinate system
of this class has associated four families of null 3-surfaces or a
null coframe, whose mutual intersections give six families of
spacelike 2-surfaces and four congruences of spacelike lines.

   In this primary positioning
system, an user at any event in a given spacetime region can know
its proper coordinates. The four proper times of four satellites
 $ (\{\tau^A \};\, A = 1,2,3,4$) read at an event by a receiver or user
constitute the null (or light) proper emission coordinates or user
positioning data of this event, with respect to four SVs, see Fig.
4. These four numbers can be understood as the ``distances"
between the event and the four satellites.

The 3-null surface $\tau^A = const$, for each A, represents the
set of events that read the proper time $\tau^A$ of an event of
the satellite path. The coordinate lines, given by the
intersection of three future light 3-surfaces $\tau^B=const$,
$\forall B\neq A$, are the integral curves of the frame vectors
$\partial/\partial{\tau^A}$. Every line in a light 3-surface  is
lightlike (along the generatrix) or spacelike. The conclusion is
that $\partial/\partial{\tau^A}$ cannot be timelike. Moreover, a
null 3-surface contains a unique null direction at each point.
Hence, the intersection among three light 3-surfaces can include
their null directions only if they coincide at the point, namely
if the three light 3-surfaces are tangent to each other. But, in
this case the Derrick-Coll-Morales coordinate system would be
degenerated, therefore the coordinate lines are always spacelike.

 In  a certain domain
$\Omega\subset\bm{R}^4$ of the grid of parameters $ (\{\tau^A
\};\, A = 1,2,3,4$), any user receiving continuously his emission
coordinates may know his trajectory. If the receiver has his own
clock, with proper time denoted by $\sigma$, then he can know his
trajectory, $\tau^A=\tau^A(\sigma)$, and  his four-velocity,
$u^A(\sigma)= d \tau^A/d\sigma$.

It is worth remarking that a similar positioning system could be
conceived for the whole Solar system by using four millisecond
pulsars.

 There is no spacetime asymmetry like in
the usual ECI Newtonian coframe  $(\it{t\,s\,s\,s })$ (one
timelike ``$\it{t}$" and three  spacelike ``$\it{s}$''). In
emission coordinates  obtained from a general real null coframe
  $ (\it{l\,l\,l\,l})=\{d\tau^1, d\tau^2, d\tau^3, d\tau^4\}$, which is neither orthogonal nor
   normalized, the contravariant spacetime metric is symmetric with null diagonal
  elements  and it has the general expression \cite{hehl}:
\begin{equation}\label{25}
(g^{AB})=  ({d\tau ^A} \cdot d\tau ^B) =\left(\begin{array}{cccc} 0 &g^{12}& g^{13}&g^{14}\\
g^{12}&0& g^{23}&g^{24}\\
g^{13}& g^{23}&0&g^{34}\\
g^{14}& g^{24}&g^{34}&0\end{array}\right),
\end{equation}
where $g^{AB}>0$ for $A\neq B$. Four null covectors can be
linearly dependent although none of them is proportional to
another. To ensure that the four null covectors are linearly
independent and span a 4-dimensional spacetime, it is sufficient
that $\text{det}(g^{AB})\neq0$. Finally, this metric has a
Lorentzian signature $(+,-,-,-)$ iff $\text{det}(g^{AB})<0$. The
expression (\ref{25}) of the metric is observer independent and
has six degrees of freedom. A splitting of this metric can be
considered changing from the six independent components of
$g^{AB}$ to a more convenient set, which neatly separates two
parameters depending only on the direction of the covectors
$d\tau^A$ from other four parameters depending on the length of
the covectors. A special case of normalized covectors giving a
regular tetrahedric configuration with one degree of freedom for
the four SVs was also considered in \cite{rovelli,hehl} for a
Minkowski spacetime.

 {\sl The SYPOR is not only a positioning system,
 but  it can also be made autonomous or autolocated}. Let us define
 the meaning of autonomous or autolocated. A system is
 autonomous or autolocated if any receiver
  can determine its spacetime path as well as
the trajectories of the four satellites, solely on the basis of
the information received during a proper time interval.

 Four satellites emitting, without the necessity of a
synchronization convention,  not only their proper times $ \tau
^A$, but also the proper times $ \tau^{AB}$ of  three close
satellites received by the satellite $A$ in $\tau^A$ (in total $
\{\tau^A, \tau^{AB}\};\, A \neq B; \,A, B = 1,2,3,4$), constitute
an autonomous positioning system. This is because  the three
proper times, together with  its proper time, that a satellite
clock receives from other SVs constitute its proper emission
coordinates. Note that, today, with the new generation of
satellites in operation in the GPS, this procedure of proper time
auto navigation with cross-links can be  technically fulfilled.

The sixteen data $ \{\tau^A, \tau^{AB} \} $ or emitter positioning
data received by a user, which allows the user to know his
trajectory and the trajectories of four SVs,
 constitute a local chart of the null
contravariant coordinate system of the atlas obtained by means of
the whole  constellation of SVs. Recently in \cite{coll6} some
explicit examples have been studied for the simpler
1+1-dimensional Minkowski case. If the constellation  is located
around the Earth, like in the Galileo, would be needed more than
four satellites and the redundancies could be used to model the
true gravitational field acting on the constellation.

\subsection{Positioning and gravimetry}
In General Relativity, the gravitational field is described by the
spacetime metric. If this metric is  exactly known {\it a priori},
the system just described will constitute an ideal positioning
system (and the components of the metric could be expressed in
Derrick-Coll-Morales coordinates). In practice, the true spacetime
metric (i.e., the gravitational field) is not exactly known (in
the GPS it is supposed to be the Schwarzschild one) and the
satellite system itself has to be used to infer it. This  problem
arises when a satellite system is used for both positioning and
measuring the spacetime metric. This implies that the positioning
problem and the gravimetry problem cannot be separated.

To solve this joint problem, the considered satellites should have
more than one clock: they may carry an accelerometer providing
more information on the spacetime metric (in fact on the spacetime
connection). Of course, in first approximation the satellites  are
in free-fall (i.e., follow a geodesic of the spacetime metric) and
consequently have zero acceleration. However, we are considering
here the realistic case where the acceleration is nonzero due, for
instance, to a  small drag in the high atmosphere and this is
measured by the accelerometers. Also, the satellites may have a
gradiometer, this would give additional information on the metric
(in fact, on the Riemann tensor of the spacetime). With these
informations (and perhaps some additional ones) an optimization
procedure can be developed (see \cite{tara,chus}) to obtain the
``best observational metric field" and, thus, the best
observational gravitational field acting on the constellation.

For the purposes of positioning alone, there is no need to use
navigation equations as in the GNSS Newtonian systems because the
emission coordinates are defined without any reference to an Earth
based coordinate system and this allows to achieve maximum
precision for the primary (Earth-surface independent) autolocated
positioning system. The primary autolocated positioning system
alone does not allow the users to situate themselves with respect
to the Earth's surface but it nevertheless permits every user to
situate himself with respect to the constellation of satellites or
with respect to any more conventional reference system deduced
from it and also allows two or more users to know their relative
positions.

Of course, for applications on or near the Earth's surface, the
primary emission coordinates should be attached to some
terrestrial secondary 4-dimensional Newtonian coordinate system,
like the 3-dimensional World Geodetic System reference system and
the  GPS coordinate time, as in the GPS, or for instance,  to the
3-dimensional International Terrestrial Reference System (ITRF),
which is also an  ECEF, and to the International Atomic Time
(TAI). The possibility appears for a space agency to concentrate
its interest in the primary positioning system and delegate to
Earth agencies the control of the terrestrial coordinates. The
function of an Earth agency must be to provide the coordinate
transformation between the primary (Earth-surface independent)
4-dimensional relativistic positioning system and the secondary
(Earth-surface dependent) 4-dimensional Newtonian reference
system.

In summary, the purpose of the  SYPOR project is to divide a GNSS
in two hierarchical systems: first, a primary system
(Earth-surface independent), made up of the constellation of
satellites. Its physical realization implies an Inter Satellite
Link  between neighboring satellites, a device in each satellite
to send to the Earth the links (physical local proper times)
directly received by each satellite, and a device in some of the
satellites of the system to connect the constellation of
satellites to the 3-dimensional International Celestial Reference
System (ICRS). Secondly, a secondary system (Earth-surface
dependent), coupling an intrinsically retarded Earth 3-dimensional
reference system (WGS 84 or ITRF) and a global conventional time,
as the GPS one, to the 4-dimensional immediate primary main
system.

\section{Conclusions}
The aim of this work was to offer  a simple introduction to the
relativistic effects on the current GNSS: GPS and GLONASS. These
systems are based on a Newtonian model which is corrected
numerically by important ``relativistic effects". This model uses
the ECI and ECEF ``Newtonian" reference frames, respectively for
time reference and navigation. The influence of Special and
General Relativity on the GNSS is considered through the
exposition of the main relativistic effects of time dilation and
Einstein gravitational shift. It is shown that if these effects
were not taken into account  the GNSS would fail after only a tiny
fraction of an hour. Expressed differently, the correct operation
of the  GNSS confirms Special Relativity and  the weak field and
slow motion limit of General Relativity.

 On the other hand, a constellation of SVs with satellites that exchange
 their proper time among them and with Earth users, is  a fully
relativistic system. Hence, this allows to use a contravariant
light coordinate system in a GNSS   that does not exist in a
Newtonian spacetime and so an autonomous positioning system can be
conceived, as in the SYPOR project.

 The application of the SYPOR project, would make of the Galileo project
  a {\sl fully relativistic}
autonomous positioning system. It would not be necessary to
correct the errors that arise from the relativistic effects (one
obtains the sum of the post-Newtonian series in $c^{-n}$) and it
would have additional technological advantages. Finally, the
satellite system could be used for both, positioning and measuring
 the gravitational field acting on the constellation.

\section*{Acknowledgments}
  I am grateful to Bartolom\'e  Coll by the ideas on the SYPOR
project  for the Galileo that he communicated to me. Also I thank
 A. San Miguel, M. del Mar San Miguel and F. Vicente  for
suggestions and a critical reading of this work. Finally, I am
grateful to an anonymous referee for its comments. The author is
currently partially supported by the grants of the Spanish
Ministerio de Educaci\'on y Ciencia MEC ESP2006-01263, MCyT
TIC2003-07020 and Junta de Castilla y Le\'on VA065A05.


\begin{thebibliography}{[00]}
\bibitem{comment} We remark that in the special case of circular
orbits $dr = 0$, the metric coefficient of spatial curvature
$g_{rr}$, i.e., the prefactor of $dr^2$ in Eq. (\ref{1}), has no
influence on the result of Eq. (\ref{3}). So, in this
approximation we could start from a metric that comes from  the
Equivalence Principle (local equivalence of inertia and gravity
for experiments not sensitive to tidal forces), that reads:
$$ds^2 = c^2\,d\tau^2=
\left(1-\frac{2GM}{rc^2}\right)\,c^2\,dt^2 -
 dr^2 -r^2 d\psi^2~,
 $$
and it is not necessary to begin from the full nonlinear
Schwarzschild metric Eq. (\ref{1}) or even from its linear
approximation. To obtain the above metric  only the part of the
Einstein Equivalence Principle that relies on Local Position
Invariance (LPI) is needed. LPI states (see \cite{Will93}) that
the outcome of any local non-gravitational experiment is
independent of where and when in the universe it is performed.
\end{thebibliography}
\end{document}